\documentclass[prb,preprint]{revtex4-1}

\usepackage[draft]{hyperref} 
\usepackage{amsmath}  
\usepackage{amsfonts} 
\usepackage{graphicx} 

\begin{document}


\title{Coherence and information in a fiber interferometer}
%
\author{Agla\'e Kellerer}
\affiliation{Department of Physics, University of Cambridge, UK}
\email{ak935@cam.ac.uk} 

\author{Sidney Wright}
\affiliation{Department of Physics, University of Cambridge, UK\\
Centre for Cold Matter, Blackett Laboratory, Imperial College London, UK}
\email{s.wright15@imperial.ac.uk}

\author{Sylvestre Lacour}
\affiliation{Department of Physics, University of Cambridge, UK\\ Observatoire de Paris-Meudon, France}
\email{sylvestre.lacour@obspm.fr}

\date{\today}

\begin{abstract}

We present an experiment based on a fibered Mach-Zehnder interferometer. The aim is to familiarize students with fibered optics and interferometry, and to improve their understanding of optical amplification. The laboratory project has two parts:
in a first part, the students modulate the optical path of the interferometer to study the spectra of light sources via Fourier Transform Spectroscopy. In a second part, an optical amplifier is placed in one or both arms of the interferometer.
The set-up uses monomode, polarization-maintaining fibers that propagate light of 1.5\,$\mu$m wavelength. In this article, we describe the set-up and the analysis of the measurements, and we present results from student reports. All components are part of standard optical catalogues. Even though the experiment is based on fibered optics, it is robust to manipulation 
(it is however relatively expensive $\sim \pounds 15\,000$): We describe our efforts to protect the components from damage. 
 This experiment is now offered as a 2-week project for third-year Physics students. 
The experiment may likewise be of value in early graduate level laboratory courses. 

\end{abstract}

\maketitle

\section{Introduction}

During the International Year of Light we have developed a fiber-based optical interferometer for undergraduate lab courses. 
The experiment aims to familiarize the students with interferometry, Fourier Transform spectroscopy and optical amplification.

The set-up is based on a fibered Mach-Zehnder interferometer. This type of interferometer is classically used in  telecommunication  to control the amplitude of  optical signals.\cite{g0,g1,g2,g3,g4,g5,g6,g7} The first part of the laboratory project is concerned with Fourier Transform Spectroscopy (FTS). An advantage of the FTS is that it needs only one detector pixel. When infrared detector arrays were not commonly accessible, FTS was the preferred method for spectroscopy in astronomy. The students use two sources (a Fabry-Perot laser and a superluminescent diode), they modulate the Optical Path Difference (OPD) between the two arms of the interferometer, and Fourier transform the amplitude of the fringes as a function of OPD. From this they deduce the spectrum of the light source. 

In a second part of the project, an optical amplifier is placed in one and then both arms of the interferometer.  The pattern observed behind an interferometer shows interference fringes, provided there is no way to determine which arm a photon has passed. 
When coherent photon pulses from a laser pass an optical amplifier, they stimulate the emission of identical, coherent  photons by the atoms of the amplifier medium. 
The student is invited to predict the outcome of this experiment. 
Are the  stimulated  photons  localized in one arm?  In this case, they should not contribute interferences. Or does the process of stimulated emission  not constrain the position of the incoming photons, so that the interference pattern is preserved and even amplified through the contribution of the stimulated emissions? The student will observe that the amplitude of the fringes increases with the amplifier gain, while the normalized contrast decreases. Thus the incoming and  stimulated photons contribute both to interferences. 
The effect of an amplification of gain $g$ is equivalent to enlarging one of the holes in a classic Young  experiment by a factor $g=A_2/A_1$, where $A_1$ and $A_2$ are the areas of the holes. The students also understand that spontaneously emitted photons  add a continuous incoherent signal, which decreases the contrast of the fringes. 

Our experiment is based on fibered optics, which eliminates the need for  optical alignment -- rather than aligning an optical component, one connects a fiber. It also exposes students to fiber technology. All components are off-the-shelf, i.e. from standard optical catalogues. 
Section\,\ref{sec:inst} presents the design and set-up of the instrument, section\,\ref{sec:report} details the different parts of the laboratory project and presents results from student reports. 

The American Journal of Physics has published several articles on Mach-Zehnder interferometers built from bulk optics.\cite{mz1,mz2,mz3,mz4,mz5} Other articles have described fibered Fabry-Perot or Michelson interferometers as instructional tools.\cite{fib1, fib2, fib3} The experiment that we set up -- a fibered interferometer used in combination with optical amplifiers -- is commonly used to  realize high-speed optical switches and logical gates.\cite{g0,g1,g2,g3,g4,g5,g6,g7} To our knowledge its use for educational purposes has not yet been described.

\section{Experimental set-up}\label{sec:inst}

\begin{figure}[htbp]
\begin{center}
\includegraphics[width=\textwidth]{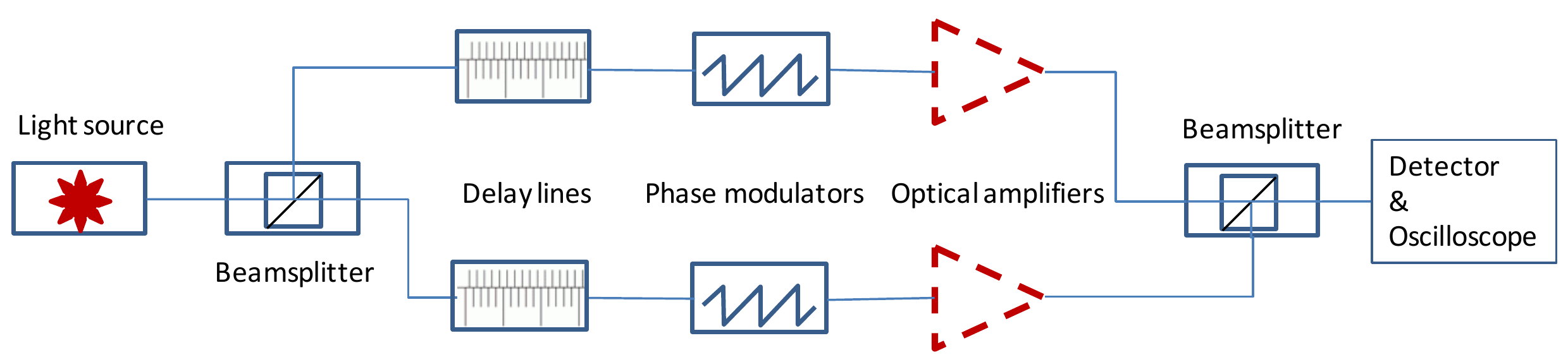}
\caption{ Setup of the Mach-Zehnder interferometer: In addition to the delay lines, phase modulators allow to visualize an interference pattern on an oscilloscope. An optical amplifier is later added in one, and then both arms of the interferometer. All optical components are fibered. }
\label{fig:setup}
\end{center}
\end{figure}

\begin{figure}[htbp]
\begin{center}
\includegraphics[width=\textwidth]{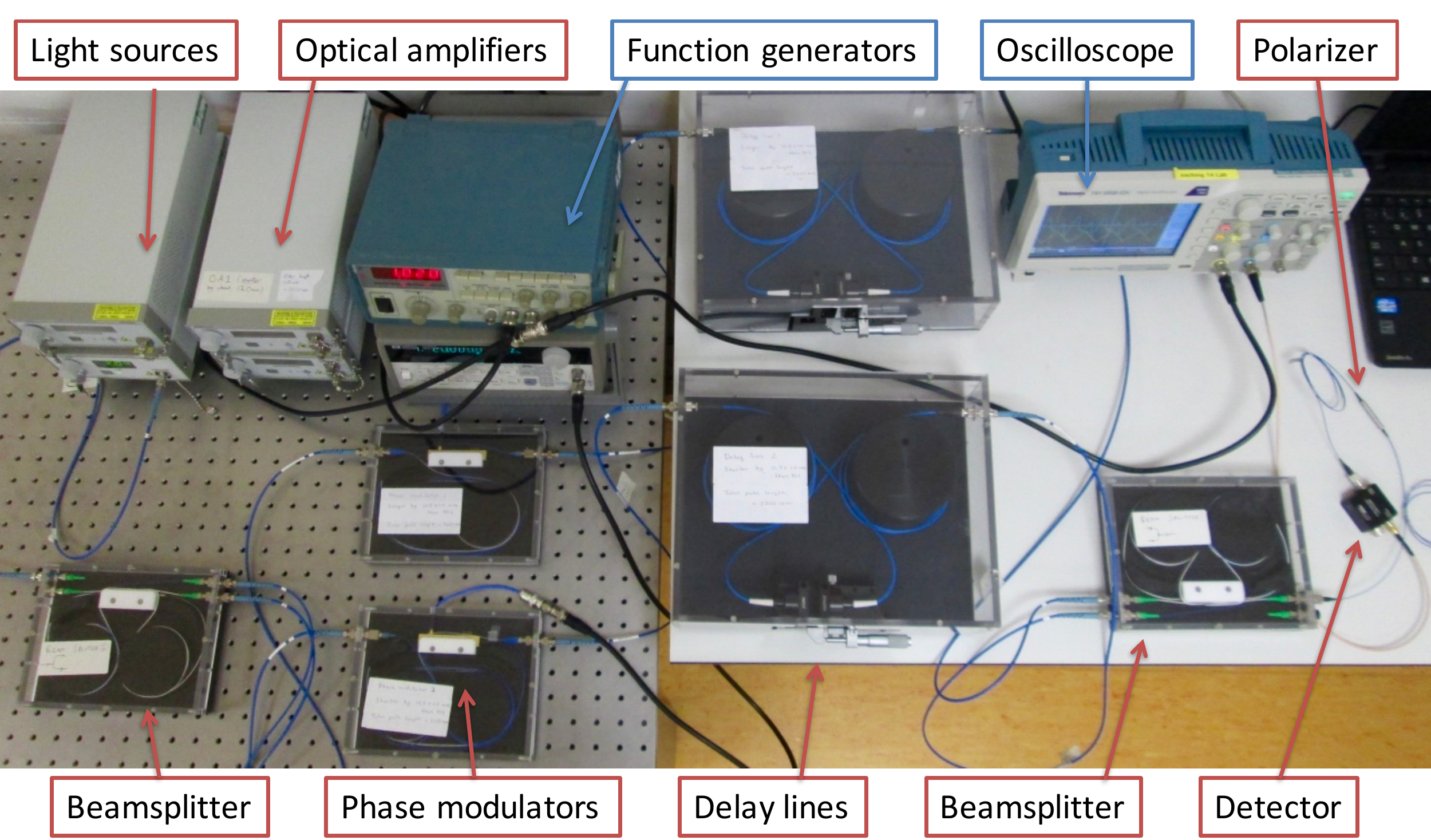}
\caption{ Picture of the set-up. The components in red are fibered. The protective plexiglas boxes were manufactured by our workshop. Part of the experiment is here placed on an optical table, but this is not a necessity. }
\label{fig:photo}
\end{center}
\end{figure}

The set-up of our Mach-Zehnder interferometer is shown on Fig.\,\ref{fig:setup}: 
The light is emitted either by a narrow-band Fabry-Perot source or a wider-band super-luminescent diode (SLD). Both sources are centered at $\lambda=1.55\,\mu$m. We chose to work at this telecommunication wavelength, because it allows us to use a broad range of components available from standard optics catalogues. Table\,\ref{tab:comp} lists the components used in this experiment. 
Figure\,\ref{fig:photo} shows a picture of the set-up.

The Fabry-Perot  source has several thin spectral lines and  fringes are obtained even far from path-length equality.  The path lengths are  equalized by adjusting the delay lines and by following the gradient of increasing fringe contrast. Afterwards, the SLD is used to measure fringe contrasts because its fringe pattern is more stable: with the Fabry-Perot laser, reflections on optical interfaces easily produce spurious interference effects and this generates instabilities in the fringe pattern.

The light is propagated along monomode, panda-style polarization-maintaining fibers (Nufern PM1550-XP). It is important that the fibers are polarization maintaining so that the light field doesn't fluctuate between non-interfering polarization states. 
All other components are likewise polarization-maintaining, and therefore bi-refringent. 
We have  bought a fibered polarizer to remove any remaining crosstalk between polarizations. 
The students may place the polarizer either before or after the interferometer.

The light is sent along the two arms of the interferometer via a beam-splitter. 
A Mach-Zehnder set-up was chosen because it provides ample space for optical components in the arms. The path lengths of the two arms are equalized via delay lines. These lines are the only components that were not purchased from Thorlabs. They are however available as standard catalogue items from  OzOptics. Each line introduces a maximum delay of 2.5\,cm, thus the total amplitude for path-length equalization is  5.0\,cm when a delay-line is placed in each arm. The path equalization is done manually via a micrometer screw. The experiment could likewise be designed with motorized delay-lines and different delay amplitudes.  

The optical path-difference is rapidly modulated with a Lithium Niobate phase-modulator. 
When the  modulator is addressed with a sawtooth or a triangular signal of roughly $10\,$V amplitude, the phase changes linearly with an amplitude of approximately $4\,\pi$. The frequency of the signal typically lies around a few hundred Hertz. The modulator allows to visualize a fringe pattern on the oscilloscope.
It has an SMP connector, thus a BNC to SMP cable is required to connect a function generator to the modulator. The cable was  made by our  kind laboratory technician. 
We place a modulator in each arm and address them with inverted signals. The amplitude of the phase modulation is then doubled and more fringes are visible on the oscilloscope. 
In order to address the two modulators with inverted signals, the two function generators need to be driven synchronously and in anti-phase. This is done by using the `sync' output of one generator as a trigger signal for the other generator, i.e. by connecting the  `sync' output to the `external trigger' input of the other generator. Alternatively, one may use a function generator that directly outputs two anti-phased signals. 

In the second part of the experiment an optical amplifier is placed in one and then both arms. We use Erbium Doped Fiber Amplifiers (EDFA), which amplify light around 1.55\,$\mu$m.
The light is  recombined via a second beam-splitter and sent onto a single-pixel detector. The detector generates a voltage that is proportional to the incident light intensity. This voltage is visualized on an oscilloscope. The oscilloscope data can be stored in .csv files. 

Initially we bought 1\,m long monomode fibers to link the different components of the interferometer. However, all components have slightly different path lengths: the path lengths of the two phase modulators differ by 1\,cm, while the path lengths of the amplifiers differ by 12\,cm. The amplitude of the delay lines (5\,cm) was not sufficient to account for those differences. We therefore bought fibers of different lengths, between 1.0\,m and 1.3\,m in increments of 0.1\,m. These fibers are not part of the standard Thorlabs catalogue, but  can be ordered from their website.\cite{Thorlabs}

Almost all the optical components have FC-APC connectors (Fiber Connection-Angle Physical Contact): The fibers are cut at $8^\circ$ and reflections are then negligible. 
This is in contrast to FC-PC connectors (Fiber Connection-Physical Contact) where fibers are cut at right angle and hence, if a fiber is left unconnected, light is reflected with a Fresnel reflection coefficient $R=4\%$. 
The Fabry-Perot laser and the phase modulators were however only available with FC/PC connectors. We thus purchased patch cables that convert FC-PC to FC-APC (Thorlabs part number P5-1550PM-FC-2). 

\begin{table}
\begin{tabular}[c]{l | l | l}  
Description & Quantity  & Part number \\ \hline 
Fabry-Perot laser & 1 & S1FC1550PM \\
Superluminescent diode & 1 & S5FC1005P \\
Beam splitter & 2 & PC1550-50-APC  \\
Delay lines & 2 & ODL-200-11-1550-8/125-P-40-3S3S-3-1  \\
Phase modulator & 2 & LN65S-FC  \\
Optical amplifier & 2 & S9FC1004P \\
Polarizer & 1& ILP1550PM-APC \\
Detector & 1 & DET01CFC/M  \\
1\,m fibers & $\sim 15$ & P3-1550PM-FC-1 \\
1.1,1.2,1.3\,m fibers & 2 of each & Custom ordered \\
FC/PC to FC/APC patch cables & 5 & P5-1550PM-FC-2 \\ 
Fiber connectors & $\sim 15$ & ADAFCPM2 \\
\end{tabular}
\label{tab:comp}
\caption{Purchased parts. All parts are manufactured by Thorlabs except for the delay lines from OzOptics. }
\end{table}

Fiber optical components are fragile. The fibers may be broken if they are twisted or strongly bent. The fiber interfaces can also easily be damaged and students are reminded to systematically protect the interfaces with  plastic covers. 
The machine shop built protective plexiglas boxes for the beam-splitters, the phase modulators and the delay lines. 
These are transparent and their cover can be unscrewed, if the students wish to measure the fiber lengths. Note that we have written the approximate path-length of each component on its box.  Hence, the students mostly leave the components untouched, which  reduces the risk of damaging the most expensive optical components. 
The light sources, amplifiers and the detector are delivered within boxes. The different boxes are then connected via  monomode fibers that have protective outer jackets (3\,mm in diameter).

The two laser sources and the optical amplifiers are class 1M lasers. These lasers are safe and the maximum permissible radiation exposure can not generally be exceeded. However, the students should still be warned never to look  onto a fiber output directly, especially as radiation at 1.5\,$\mu$m is not visible. 

\section{Measurements}\label{sec:report}

The project is  divided into two parts: the students first use the delay lines of the Mach-Zehnder to do Fourier Transform Spectroscopy (FTS).  In this first section the contrast as a function of optical path-length difference is used to derive the spectra of different light sources.
In the second part of the project an amplifier is placed in one and then in each arm of the interferometer. 
For a shorter project the experiment can be reduced to its first part. 

\subsection{Fourier Transform Spectroscopy}

FTS is most easily explained by considering the wave nature of light. The electric field is written as a superposition of monochromatic waves. It is represented as a complex function of frequency $\nu$ and time $t$:
 \begin{equation}
 E(t)=\int_{-\infty}^{\infty} a(\nu) \exp(- j 2 \pi \nu t)\, \mathrm{d}\nu \, ,
 \label{eq:E}
 \end{equation}
We consider frequencies between  ${-\infty}$ to ${+\infty}$ to simplify the Fourier transformations. The amplitude of negative frequency waves can be set to zero. $a(\nu)$ is a complex parameter that contains the phase and amplitude of each monochromatic wave. Note that we would use a vector instead of a scalar if we wanted to account for polarization. Here, we neglect the effect of polarization, since we are using polarization maintaining fibers. 
The intensity observed at a time $t$, on a hypothetical detector of infinitively small response time, is related to the electric field via: $I(t)=\Re[E(t)]^2$. 
  
The detector used in this experiment is a photodiode with a 1\,GHz bandwidth. The signal is thus integrated over 1\,ns, which is  long compared to the frequency of the field: $\nu = 200 \,$THz at $\lambda=1.5\,\mu$m. We can therefore make the following approximation:
 \begin{equation}
I=< \Re[E(t)]^2 > =  < \Im[E(t)]^2 >= \frac{< \Re[E(t)]^2 > +  < \Im[E(t)]^2 > }{2} = \frac{< E(t) E^*(t)>}{2} 
\label{eq:EE}
 \end{equation}
where $<>$ is the average over the integration time of the detector. The above relations use the fact that the squares of  cosine and sine have the same average over many periods.

According to Eq.~(\ref{eq:E}), $E(t)={\rm FT\/}[a(\nu)]$. 
Hence, 
\begin{eqnarray}
I&=& \frac{1}{2} <\int_{-\infty}^{\infty}  a(\nu) \exp(- j 2 \pi \nu t)\,\mathrm{d}\nu \, \int_{-\infty}^{\infty}  a^*(\nu') \exp( j 2 \pi \nu' t)\,\mathrm{d}\nu' >
 \end{eqnarray}

The integration times are long compared to the wave periods, hence:
 \begin{equation}
 < \exp( j 2 \pi \nu t- j 2 \pi \nu' t) > = \delta (\nu-\nu')
 \label{eq:dirac}
 \end{equation}
where $\delta (\nu-\nu')$ is the dirac function which is null for any value of $\nu$ different from $\nu'$. Thus
 \begin{equation}
 < a(\nu) \exp(- j 2 \pi \nu t)\,\mathrm{d}\nu \cdot a^*(\nu') \exp( j 2 \pi \nu' t)\,\mathrm{d}\nu' > = a(\nu) a^*(\nu') \cdot \delta (\nu-\nu') \,\mathrm{d}\nu \, \mathrm{d}\nu' 
 \end{equation}
and
 \begin{equation}
I= \frac{1}{2}  \int_{-\infty}^{\infty}  a(\nu) a^*(\nu) \, \mathrm{d}\nu=\frac{1}{2}  \int_{-\infty}^{\infty}  |a(\nu)|^2 \, \mathrm{d}\nu\,.
 \end{equation}
This relation means that the polychromatic intensity is the sum of the intensities contained in the monochromatic waves. This is re-assuring for energy conservation, but it's not trivial for the students to demonstrate this. 

The interferometer is set up as shown on Fig.\,\ref{fig:setup}, without optical amplifiers and with the Fabry-Perot laser source. 
The phase modulators are controlled with triangular signals of $\sim 10\,$V amplitude at a few hundred Hertz. The modulators are operated in anti-phase. The path lengths in the two arms should be roughly equalized, with a precision of a few centimeters, so that the final adjustment can be made with the micrometer screws of the delay lines. The spectrum of the Fabry-Perot source contains narrow lines and fringes are therefore  observed even far from  path-length equality: Fringes may be observed a few meters away from path-equality.
The paths are equalized by adjusting the delay lines and by following the gradient of increasing fringe contrast.  Once the path lengths are roughly equal, fringes can also be found with the SLD source. 

The evolution of the interference pattern as a function of optical path-length difference is then used to characterize the spectral response of different light sources. The intensity observed on the detector results now from the interference of two waves that have travelled different optical path-lengths. This difference translates into a time delay  between the two waves coming from both arms of the interferometer:
 \begin{eqnarray}
E_{\rm 1}(t)&=& K_1 \int_{-\infty}^{\infty} a(\nu) \exp(- j 2 \pi \nu t)\, \mathrm{d}\nu \\
E_{\rm 2}(t)&= & K_2 \int_{-\infty}^{\infty} a(\nu) \exp(- j 2 \pi \nu (t-\delta))\, \mathrm{d}\nu
 \end{eqnarray}
 where $K_1$ and $K_2$ are real values which correspond to the splitting ratio of the light wave in the two arms. The light intensities in arms 1 and 2 are $I_1=|K_1|^2 I_0$ and $I_2=|K_2|^2 I_0$, where $I_0= 0.5  \int_{-\infty}^{\infty}  |a(\nu)|^2 \, \mathrm{d}\nu$. For flux conservation, $|K_1|^2+|K_2|^2=1$ if the splitting is lossless. 
 The time delay is $\delta=OPD/c$. We assume that the beam-splitter is achromatic so that both fields have the same spectral dependance $a(\nu)$. The  electric field at the output of the interferometer is the sum of the two fields:
 \begin{eqnarray}
E_{\rm MZ}(t)&=&K_1  \int_{-\infty}^{\infty} a(\nu) \exp(- j 2 \pi \nu t)\, \mathrm{d}\nu +  K_2 \int_{-\infty}^{\infty} a(\nu) \exp(- j 2 \pi \nu (t-\delta))\, \mathrm{d}\nu \nonumber \\
&=& K_1 \, {\rm FT\/}[a(\nu)] + K_2 \, {\rm FT\/}[a(\nu) \exp( j 2 \pi \nu \delta)] 
 \end{eqnarray}
Thus, 
 \begin{eqnarray}
 < E_{\rm MZ}(t) E_{\rm MZ}^*(t)> &=&  |K_1|^2 <{\rm FT\/}[a(\nu)] \cdot {\rm FT\/}^*[ a(\nu)]  > \nonumber \\ 
 && + |K_2|^2 <{\rm FT\/}[a(\nu) \exp( j 2 \pi \nu \delta)] \cdot {\rm FT\/}^*[ a(\nu) \exp( j 2 \pi \nu \delta) ]  >  \nonumber \\ 
&&  + K_1K_2  <{\rm FT\/}[a(\nu)] \cdot {\rm FT\/}^*[ a(\nu) \exp( j 2 \pi \nu \delta) ]  > \nonumber \\ 
&&  + K_1K_2 <{\rm FT\/}[a(\nu) \exp( j 2 \pi \nu \delta)]  \cdot {\rm FT\/}^*[ a(\nu)] > 
 \end{eqnarray}
 The four terms simplify because the detector integration-times are long compared to the wave frequency (see Eq.\,\ref{eq:dirac}):
 \begin{eqnarray}
<{\rm FT\/}[a(\nu)] \cdot{\rm FT\/}^*[ a(\nu)]  > &= & \int_{-\infty}^{\infty}  |a(\nu)|^2 \, \mathrm{d}\nu \\
 <{\rm FT\/}[a(\nu) \exp( j 2 \pi \nu \delta)] \cdot {\rm FT\/}^*[ a(\nu) \exp( j 2 \pi \nu \delta) ]  > &=& \int_{-\infty}^{\infty}  |a(\nu)|^2\, \mathrm{d}\nu \\
<{\rm FT\/}[a(\nu)] \cdot {\rm FT\/}^*[ a(\nu) \exp( j 2 \pi \nu \delta) ]  > &=& \int_{-\infty}^{\infty}  |a(\nu)|^2 \exp(- j 2 \pi \nu \delta) \, \mathrm{d}\nu \\
<{\rm FT\/}[a(\nu) \exp( j 2 \pi \nu \delta)] \cdot {\rm FT\/}^*[ a(\nu)] >   &=& \int_{-\infty}^{\infty}  |a(\nu)|^2 \exp( j 2 \pi \nu \delta) \, \mathrm{d}\nu 
 \end{eqnarray}
 Hence, since we have established in Eq.~(\ref{eq:EE}) that $ I(\delta) = \frac{1}{2} < E_{\rm MZ}(t) E_{\rm MZ}^*(t)> $:
 \begin{equation}
 I(\delta) = \frac{|K_1|^2+|K_2|^2}{2}\int_{-\infty}^{\infty}  |a(\nu)|^2 \, \mathrm{d}\nu +  \Re\left[K_1K_2\int_{-\infty}^{\infty}  |a(\nu)|^2 \, \exp( -j 2 \pi \nu \delta)  \,\mathrm{d}\nu \right]\,.
  \end{equation}
This last equation simplifies when using the intensities in arms one and two : \\ $I_1=  0.5\, |K_1|^2 \, \int_{-\infty}^{\infty}  |a(\nu)|^2 \, \mathrm{d}\nu$ and
$I_2=   0.5\,|K_2|^2 \, \int_{-\infty}^{\infty}  |a(\nu)|^2 \, \mathrm{d}\nu$, as well as the normalized spectral density as defined by:
 \begin{equation}
  S(\nu)=
\frac{|a(\nu)|^2 + |a(-\nu)|^2}{\int_{-\infty}^{\infty} 2 |a(\nu)|^2 \, \mathrm{d}\nu} 
\label{eq:S}
 \end{equation}

This yields:
 \begin{eqnarray}
 \nonumber
 \Re\left[\int_{-\infty}^{\infty}  |a(\nu)|^2 \, \exp( -j 2 \pi \nu \delta)  \,\mathrm{d}\nu \right]
 &=&\int_{-\infty}^{\infty}  |a(\nu)|^2 \frac{\exp( j 2 \pi \nu \delta) + \exp( -j 2 \pi \nu \delta) }{2} \,\mathrm{d}\nu \\ \nonumber
 &=&\int_{-\infty}^{\infty}  \frac{|a(\nu)|^2 + |a(-\nu)|^2}{2} \, \exp( -j2 \pi \nu \delta)  \,\mathrm{d}\nu\\
 &=&\int_{-\infty}^{\infty}  |a(\nu)|^2 \, \mathrm{d}\nu \cdot \int_{-\infty}^{\infty}  S(\nu) \, \exp( -j2 \pi \nu \delta)  \,\mathrm{d}\nu\,.
  \end{eqnarray}
Hence:
 \begin{equation}
 I(\delta) = I_1+I_2 + 2 \sqrt{I_1 I_2} \int_{-\infty}^{\infty}  S(\nu) \, \exp( -j 2 \pi \nu \delta)  \,\mathrm{d}\nu \,.
 \label{eq:I}
  \end{equation}
One can thus relate the intensity observed at a given delay to the Fourier transform of the spectrum as defined in Eq.~(\ref{eq:S}):
 \begin{equation}
 I(\delta) = I_1+I_2 +  2 \sqrt{I_1 I_2} \cdot {\rm FT\/}[S(\nu)]\,.
 \label{eq:If}
  \end{equation}

Hence, the students can use the intensity observed at the output of the MZ to calculate the spectrum of the light source. The most straightforward approach would be to measure the intensity as a function of $\delta$, and use the relation: $S(\nu) = {\rm FT\/}^{-1}[ (I(\delta) - I_1-I_2) / 2\sqrt{I_1I_2} ]$. However, the set-up is not made to allow fast scanning over several centimeters with the delay lines. 

Instead the students use a small phase modulation, $\delta'$, obtained by means of the LiNbO$_3$  modulators, to estimate the contrast at a given path difference. The path difference is set with the delay lines.
The contrast determines the amplitude of the Fourier transform at a given $\delta$: 
 \begin{equation}
C (\delta)=\frac{I(\delta+\delta')_{\rm max\/}-I(\delta+\delta')_{\rm min\/}}{I(\delta+\delta')_{\rm max\/}+I(\delta+\delta')_{\rm min\/}}
\label{eq:C}
 \end{equation}
 where  $I(\delta')_{\rm max\/}$ and $I(\delta')_{\rm min\/}$ are the maximum and minimum intensities, as defined  in Eq.~(\ref{eq:I}), over one modulation-period of the LiNbO$_3$ phase modulators. From this, one derives: 
 \begin{equation}
C (\delta)= \left | \frac{\sqrt{I_1I_2} \, {\rm FT\/}[S(\nu-\nu_0)] }{I_1+I_2}\right | \,,
 \end{equation}
 where $\nu_0$ is the modulation frequency of the LiNbO$_3$ devices. The modulus translates the fact that the setup is not phase referenced: 
one may measure the amplitude of the fringes, but not their phase. Standard FTS systems  measure a complex visibility, which directly corresponds to the complex Fourier transform of $S(\nu)$. Here we only determine the contrast, $C (\delta)$, which is related to the auto-correlation of the light spectrum:
 \begin{equation}
C (\delta)^2 =  \frac{{I_1I_2} \, \left | {\rm FT\/}[S(\nu-\nu_0)]  \right |^2 }{(I_1+I_2)^2} =\frac{{I_1I_2} }{(I_1+I_2)^2} \cdot  {\rm FT\/}[S(\nu-\nu_0) \oplus S(\nu-\nu_0) ] \,.
\label{eq:spectrum}
 \end{equation}
 The sign $\oplus$ denotes the correlation function:
 \begin{equation}
S(\nu) \oplus S(\nu) = \int_{-\infty}^{\infty}  S(\nu') * S(\nu-\nu')   \,\mathrm{d}\nu'\,. 
 \end{equation}

\begin{figure}[htbp]
 \begin{center}
\includegraphics[width=.45\textwidth]{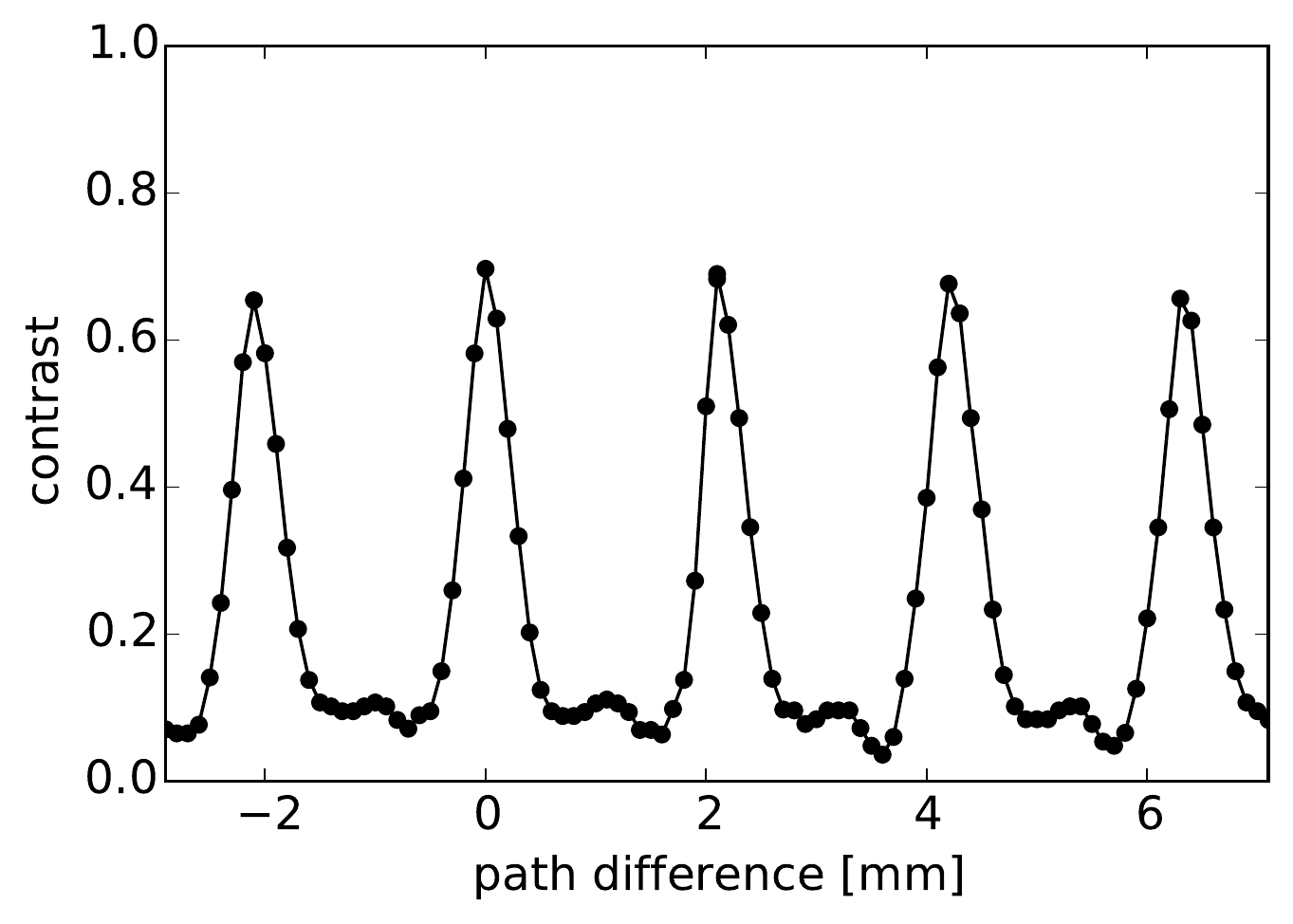}
\caption{ Fringe contrast as a function of path-difference when the light from the Fabry-Perot laser is sent through the fibered Mach-Zehnder interferometer.}
\label{fig:contrast-pathdelay}
\end{center}
\end{figure}
\begin{figure}[htbp]
 \begin{center}
\includegraphics[width=.45\textwidth]{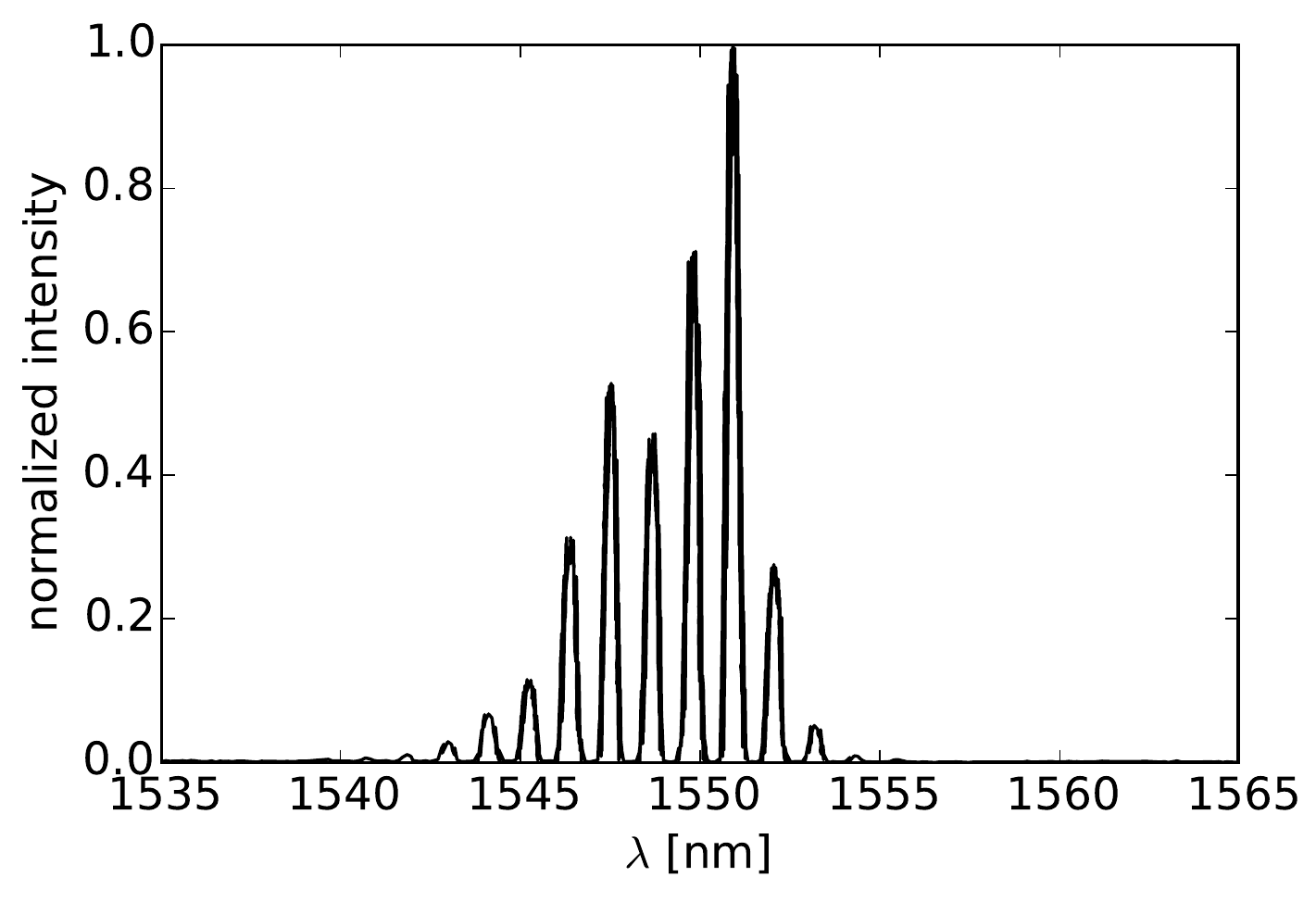}
\includegraphics[width=.45\textwidth]{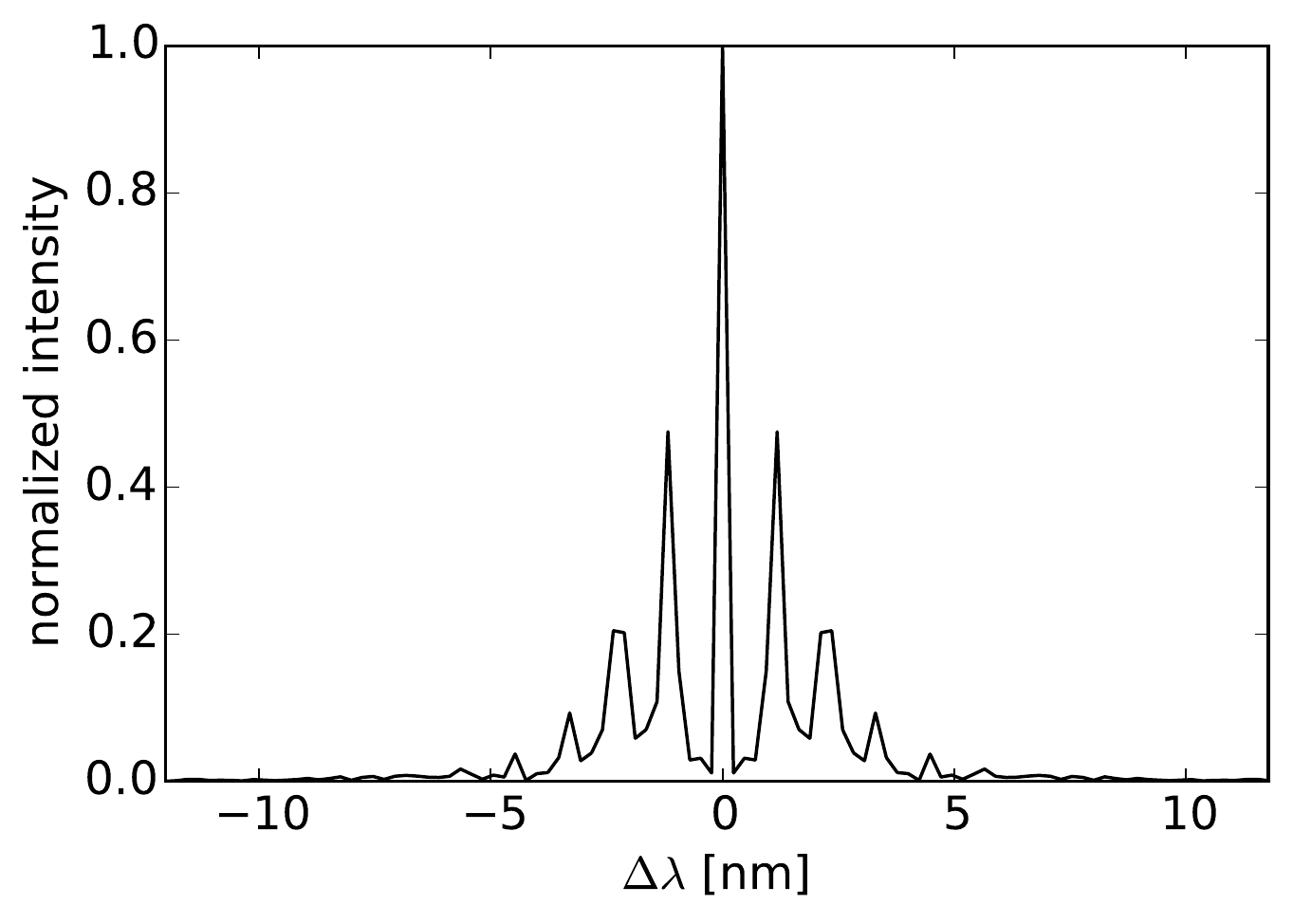}
\caption{ The spectrum specified by the manufacturer (left panel) is compared to the spectrum obtained through Fourier Transform Spectroscopy (right panel). Because the contrasts are real values, the Fourier transform of the contrast curve is centro-symmetric. The students are asked to reflect on this; the explanation is given by Eq.\,(\ref{eq:spectrum}).}
\label{fig:spectrum}
\end{center}
\end{figure}

Figure\,\ref{fig:contrast-pathdelay} traces the contrast as a function of optical path difference, using the Fabry-Perot laser. The resulting spectrum is compared to the spectrum provided by Thorlabs on Fig.\,\ref{fig:spectrum}. Resonant modes are expected when the Fabry-Perot cavity length is a multiple of half the wavelength. Hence the students can deduce the cavity length from their data: $l=\lambda^2/(2\,\Delta\lambda)$ where $\Delta\lambda$ is the distance between peaks in the spectrum. In this particular case, the student obtained a cavity length of $l=1.02\pm0.03$\,mm, close to the manufacturer specifications of approximately 1.07\,mm.

The students likewise obtain spectra for the SLD source. With a quasi-Gaussian bandwidth of 50\,nm, the SLD is perfect to understand the impact of the spectrum on the coherence length (a few tens of micrometers). The students are asked to derive the coherence length. Note that this length has various definitions, which relate either to the root-mean square deviation or to the full-width at half-maximum of the Gaussian spectrum. Either definition can be used. 

\begin{figure}[htbp]
 \begin{center}
\includegraphics[width=.45\textwidth]{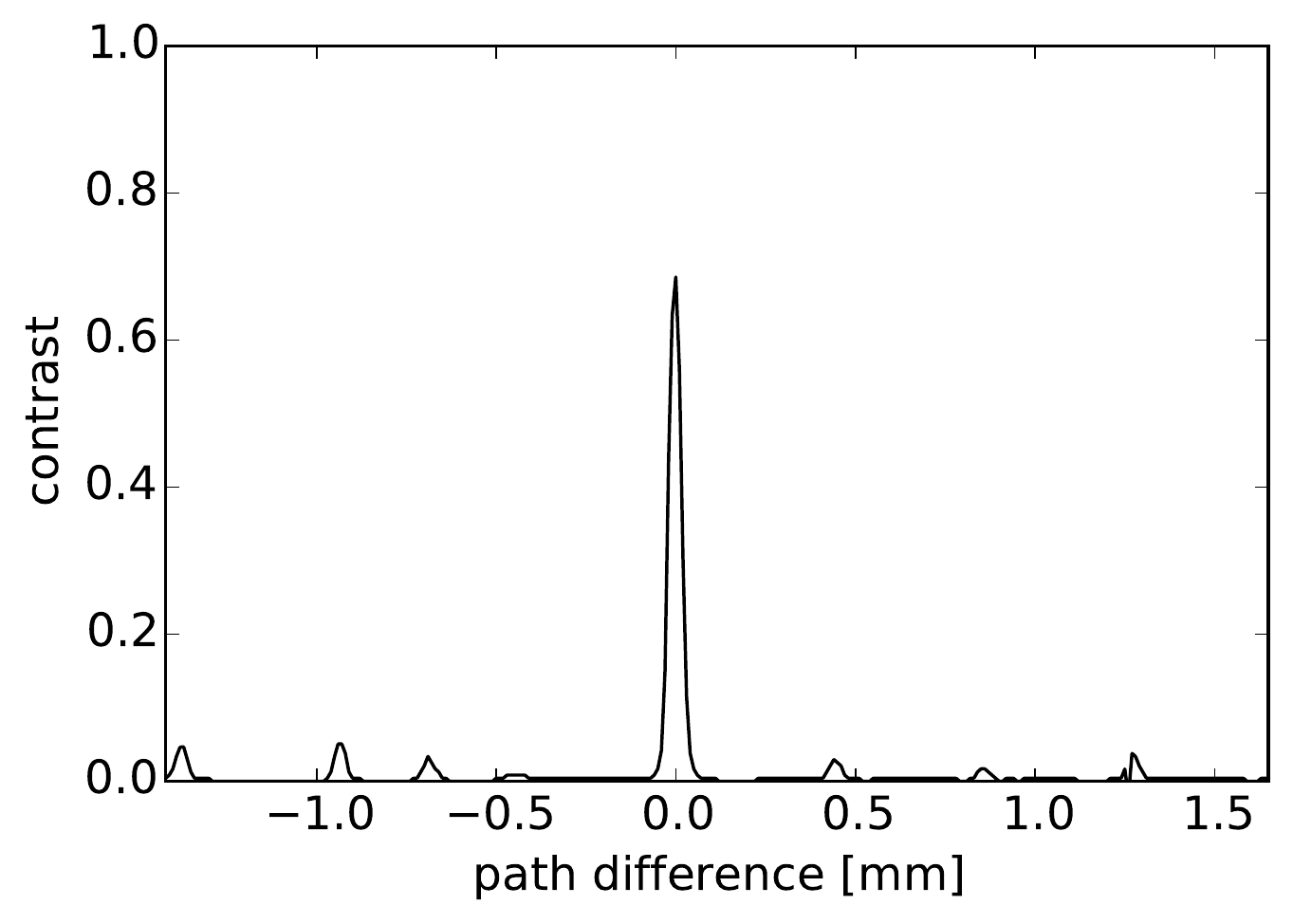}
\includegraphics[width=.45\textwidth]{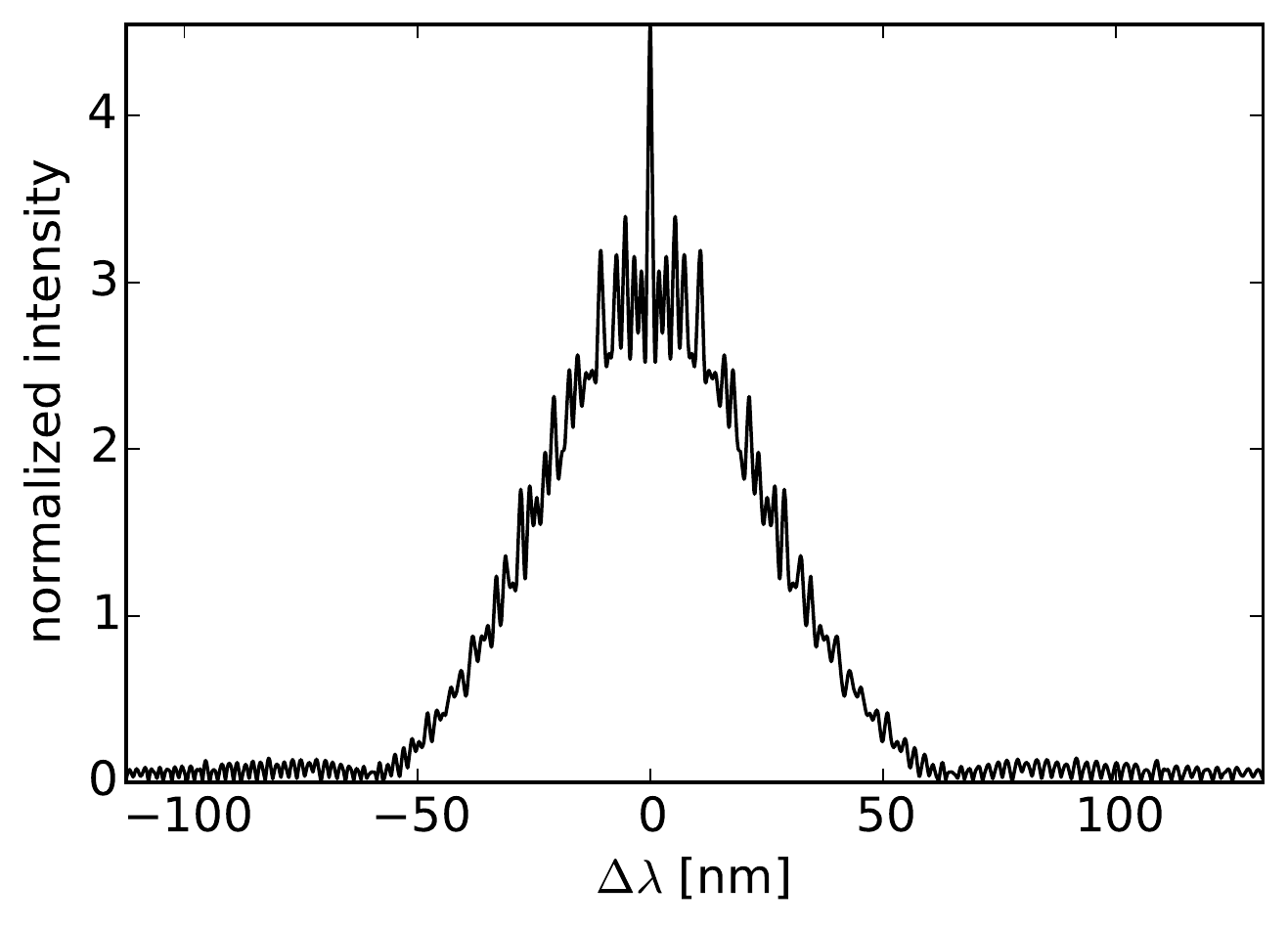}
\caption{ Left panel: Contrast as a function of path-length difference when the light from the SLD is sent through the interferometer. Right panel: The spectrum of the SLD gives a 3\,dB bandwidth of $60\pm 8$\,nm and a large scale spectral ripple wavelength of $1.8\pm0.2$\,nm -- in good agreement with the manufacturer specifications.}
\label{fig:SLD}
\end{center}
\end{figure}

Figure\,\ref{fig:SLD} shows the results from a lab report: ``The spectrum of the SLD is obtained by measuring the contrast of the fringes as a function of path length difference. The spectrum gives a 3\,dB bandwidth of $60\pm 8$\,nm and a large scale spectral ripple wavelength of $1.8\pm0.2$\,nm. These are in agreement with the manufacturer test values of 66.4\,nm and $\sim 2$\,nm respectively. 
The spectral ripple arises due to remnants of lasing effect in the SLD cavity. The cavities of the SLD have anti-reflective coating in order to prevent lasing; however, since reflection can never be completely eliminated, a slight remnant of preferred cavity modes can feature in the SLD spectrum, giving a ripple effect. The SLD cavity length is calculated as $680\pm 60\,\mu$m.''

Last, the students are asked to measure the spectra of the optical amplifier with and without a source. This is done by putting the optical amplifier at the entrance of the Mach-Zehnder interferometer. This experiment introduces the next section, where one or both optical amplifiers are placed within the interferometer.

\subsection{Amplified interference}

\begin{figure}[htbp]
\begin{center}
\includegraphics[width=0.6\textwidth]{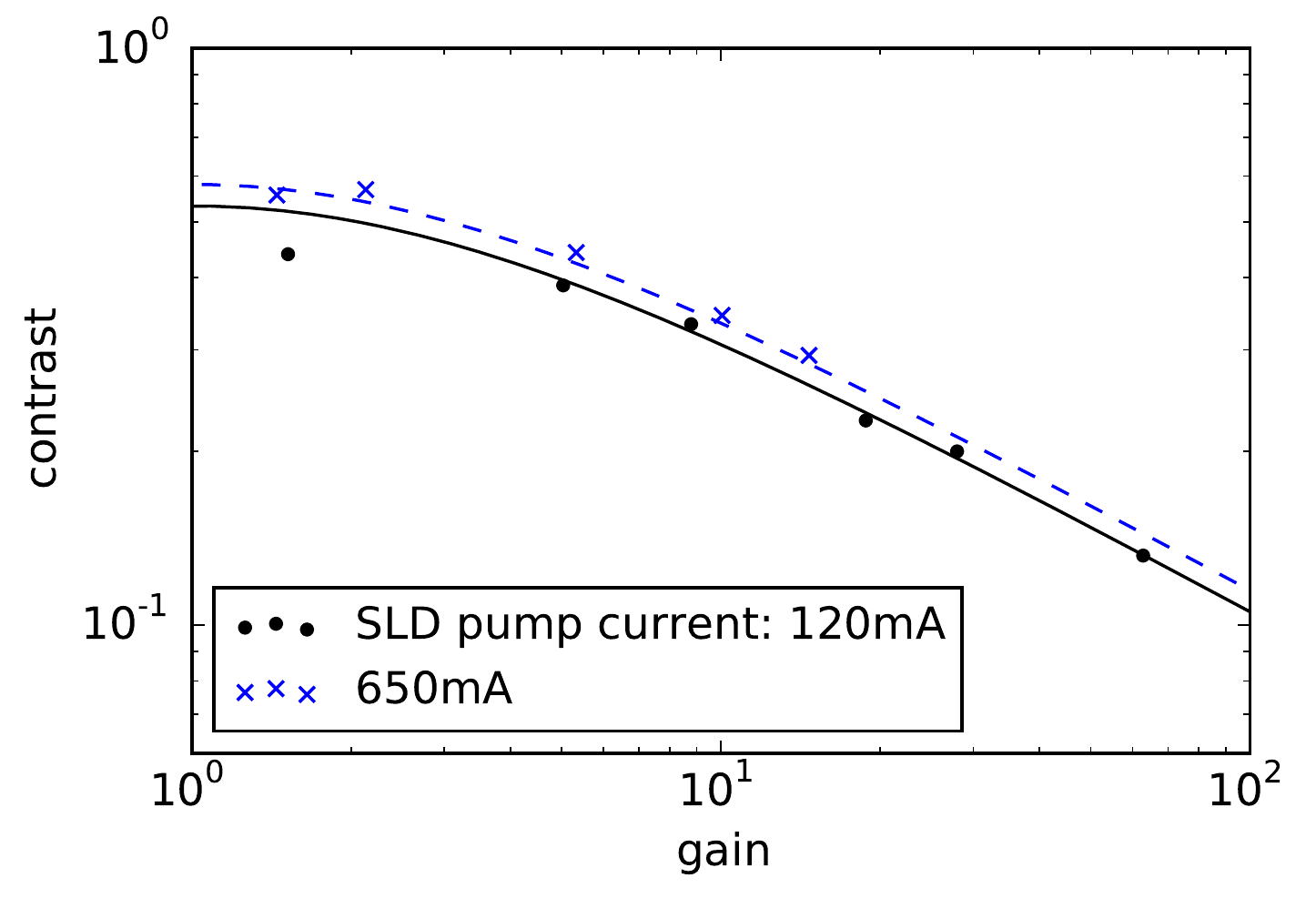}
\caption{ Logarithmic plot of the measured fringe contrast against amplifier gain. In the limit of large gain, the contrast decreases as $g^{-1/2}$, indicating that the stimulating and stimulated photons all produce interference. The two curves correspond to two different values of the source brightness.}
\label{fig:datalogplot}
\end{center}
\end{figure}    

In part two of this laboratory experiment, an optical amplifier is placed in one arm of the Mach-Zehnder interferometer. As light passes the amplifier it stimulates the emission of photons. 
The student is asked whether these photons contribute interference even though they are emitted in one arm of the interferometer. 

To answer this question, the setup of the first section is modified with the addition of an optical amplifier in arm 1. The path length of the amplifier equals approximately 3\,m. The same length of fibers thus needs to be introduced in arm 2.  

All the work performed in this part of the experiment is obtained at maximum fringe contrast, i.e. with the delay lines set  to zero path difference. For simplicity, we assume a monochromatic light source: $S(\nu)\neq0$ only if $|\nu|=\nu_0$. 
Hence, according to Eq.~(\ref{eq:If}):
 \begin{equation}
 I(\delta) = I_1+I_2 +  2 \sqrt{I_1 I_2} \cos( 2 \pi \nu_0 \delta) \,.
  \end{equation}
 or, in terms of the contrast as defined by Eq.~(\ref{eq:C}):
 \begin{equation}
 I(\delta) = I_1+I_2 +  \frac{I_1+I_2}{2} C \cos( 2 \pi \nu_0 \delta) \,.
  \end{equation}
 From this we  derive the relation between the contrast at zero OPD and the intensities in both arms:
 \begin{equation}
C = \frac{2 \sqrt{I_1 I_2}}{I_1+I_2}\,.
  \end{equation}

The aim of is to characterize the fringes obtained behind the MZ when an optical amplifier is placed in one of its arms. The intensity detected in arm 1 equals $\tilde{I_1}=I_1+I_{\rm st\/}+I_{\rm sp\/}$, where $I_1$ is the intensity that enters the amplifier, $I_{\rm st}=(g-1)I_1$ corresponds to the
 stimulated photons ($g$ is the amplifier gain) and $I_{\rm sp\/}$ corresponds to spontaneous emissions. 

If the stimulated emissions do not contribute to the fringes, then the contrast is :
\begin{eqnarray}
C_{\rm no}=\frac{2 \sqrt{I_1 I_2}}{g I_1+I_{\rm sp\/}+I_2}
\end{eqnarray}
If the stimulated emissions contribute to the fringes, then the contrast is  :
\begin{eqnarray}
C_{\rm yes}=\frac{2 \sqrt{g I_1 I_2}}{g I_1+I_{\rm sp\/}+I_2}
\end{eqnarray}

In the approximation of a large gain, $C_{\rm no}$ varies as $1/g$, while  $C_{\rm yes}$ varies  as $1/\sqrt{g}$.
Figure\,\ref{fig:datalogplot} represents the contrast as a function of gain, using the SLD at two different brightness values. A clear $1/\sqrt{g}$ dependence is approached for large amplifier gain values.

When spontaneous emissions can be neglected (i.e. for high input intensities), optical amplification and attenuation have the same effect: they change the ratio of the interfering amplitudes.  
The same evolution of contrast would also be obtained with Young holes of different size. This has been analyzed by Englert:\cite{Englert} The contrast of the fringe pattern is maximal when the knowledge of the photon position on the Young screen is minimal, i.e. when the two holes are of same size and the interferometer is symmetric. If only one hole is open, the path of the photon is perfectly known and the contrast decreases to $C = 0$. Intermediate set-ups yield contrast values $0<C< 1$.

This is not true for small input intensities, when spontaneous emissions contribute an incoherent signal, which does not produce interferences. 

Finally an amplifier is introduced in each arm of the Mach-Zehnder interferometer. 
Let $g_1$ and $g_2$ be the gain of the two amplifiers and $I_{\rm sp 1\/}, I_{\rm sp 2\/}$ the intensities from the spontaneous emissions. 
At zero path difference, the fringe contrast equals: 
\begin{eqnarray}
C=\frac{2\sqrt{g_1\,g_2\,I_1I_2}}{I_{\rm sp 1\/}+I_{\rm sp 2\/}+g_1 I_1 +g_2 I_2}
\end{eqnarray}

The intensity of the SLD laser source is fixed and the pumping currents of the amplifiers are varied. Maximum contrast is obtained when the amplifier gains, $g_1$ and $g_2$, are equal in both arms. For large gain values, when the spontaneous emissions are negligible compared to stimulated emissions, $C=2\sqrt{g_1\,g_2}/(g_1+g_2)$. The contrast approaches 1 as $g_1\sim g_2$: the fringe pattern has an excellent contrast even though most of the photons are generated inside the interferometer.

\section{Conclusion}

We have presented a new fiber-based interferometric experiment. Fiber optical components are expensive and the budget of this set-up has reached $\sim \pounds15\,000$. We feel that this investment is justified because students become familiar with widely-used fiber technology, they get hands-on experience with interferometry and improve their understanding of optical amplification. The experiment has now become part of the syllabus for third year Physics students. 

By now, five students have  successfully worked on the experiment over two-week periods (mainly over half-days, so the project could also be offered as a full-time one week project). 
The students are in their third year and they are given much liberty on the conduct of the project. The report notably asks questions on the Physics involved, rather then giving instructions on how to assemble the experiment. The students then figure out which measurements are needed to answer these questions. 
They are free to spend more or less time on the different parts of the experiment. 
We were positively surprised by the variety of the students' reports: one student spent the bigger part of the project characterizing and modeling the optical amplifiers, while another student meticulously characterized the light sources with and without amplification, via a Fourier analysis of the fringe contrast.  Their reports analyze the differences between their measurements and the manufacturer specifications. This ensures that the experiment has been well understood. 
All students managed to obtain stable fringe patterns for the various instrumental set-ups: the classic Mach-Zehnder interferometer and its modified version with an amplifier in one or both arms. Nothing has been broken yet. The experiment has thus proven reliable so far. 

Let us finally note that -- much like a report by Danan et al.\,\cite{Danan} -- the analysis of the photon trajectories inside our interferometer can come as a surprise. Intriguingly, the photons that are stimulated inside one arm of the interferometer still contribute to the interference pattern because one cannot distinguish the incoming and stimulated photons. 
Neither Danan et al.'s nor our experiment is ran in the single photon regime and the resulting interference patterns are easily understood when the wave nature of light is considered. The outcome of the experiment is, however, harder to conciliate with the particle nature of light. This intriguing evolution of the photons inside our modified interferometer helps to understand stimulated emission.

\vskip .5cm

\begin{acknowledgments}
We acknowledge our lab technician Richard King for his constant and extremely helpful assistance. We also thank John Richer and Pietro Cicuta for suggesting to develop a new experiment, for providing the necessary funds and for their general interest and support. Many thanks go to Jaan Toots, Andrei Ruskuc and Alasdair McNab for letting us use data from their lab reports. 
Finally, we are  grateful to the mechanical workshop, especially to Kevin Mott, for building the protective boxes. 
SL acknowledges grant ERC-STG-639248. 
\end{acknowledgments}

\end{document}